\documentclass[twoside,fleqn]{article}
\usepackage{espcrc2}
\usepackage[dvips]{graphics}

\title{Anisotropic Lattice and Its Application to Quark Gluon Plasma}

\author{S. Sakai\address{Faculty of Education, Yamagata University\\},
     A. Nakamura\address{Information Media Center, Hiroshima University\\}, 
and
T.Saito,\address{Department of Physics, Hiroshima University}
}
\begin{document}
\begin{abstract}
 We have studied the link-integration method for the improved
actions. With this method the $\eta$ parameter in the medium 
to strong coupling regions is obtained.   Effects of the self-energy 
terms for the $\eta$ parameters are small in the regions of $\beta$ and 
$\eta$ studied. After these investigations, the anisotropic lattice is 
used for the
calculation of transport coefficients of the quark gluon plasma.
\end{abstract}
\maketitle

\section{Link Integration For Improved Action}
\begin{figure}
\begin{center}
\scalebox{0.30}{ { \includegraphics{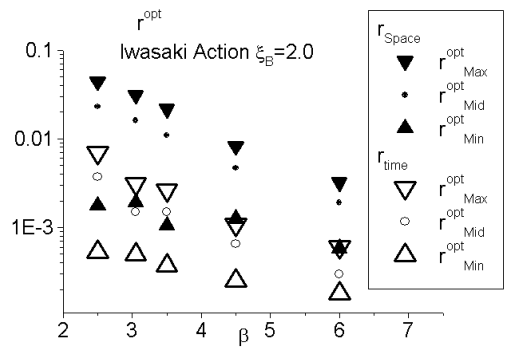} } }
\caption{An example of $r^{opt}_{max}$,
$r^{opt}_{min}$ and $r^{opt}_{mid}$ on a fully thermalized configuration 
\label{radius_vs_B} }
\end{center}
\end{figure}

If $R$ is an external source field for link variable $U$, 
the expectation value of $U$ is given by,
\begin{equation}
 <U>= 
   \frac{1}{Z} \frac{dZ(R)}{dR^{\dagger}}\label{lint1}
\end{equation}
$Z(R)$ is expressed by the modified Bessel function
$I_{1}$\cite{brower},\cite{forcrand}.
\begin{equation}
 Z(R)=\oint \frac{dx}{2 \pi i} e^{xQ} \frac{1}{z} I_{1}(2z)
\end{equation}
where
$$ z=\left(\frac{P(x)}{x}\right)^{\frac{1}{2}} $$
$$ Q=2Re(det(R))$$
\begin{equation}
\begin{array}{ll}
  P(x)= 1+xTr(R R^{\dagger}) \\
\hspace*{1cm} +\frac{1}{2} x^2 \left[(Tr(R R^{\dagger}))^2 
              - Tr((R R^{\dagger})^2)\right]\\
\hspace*{1cm}   +x^3 det(R R^{\dagger}) 
\end{array}
\end{equation}
Similarly $dZ(R)/dR$ is
written by the $I_{1}$ and $I_{2}$\cite{brower},\cite{forcrand}.\\
\indent
The path of the integration is a closed circle on the complex
plane $x$. In principle its radius $r$ is arbitrary, but numerical
integration requires adequate radius.
We apply Simpson method for the
numerical integration, and search for the region of $r$ and number of
the division $N$ where $<U>$ is stable with the change of $r$ and
$N$.
It is observed that the $<U>$ depends strongly on $r$ and $N$. 
For example if we take $N=50$, there appear spurious plateaus,
which disappears with increasing $N$. However there is a region of $r$
in which $<U>$ is stable for the change of $r$ and $N$. It 
becomes
a little wider as $N$ is increased\footnote{These features
are seen not only in the asymptotic
expansion of the modified Bessel
functions but also if
Taylor expansions of them are applied.
}. 
This region of $r$ is called an \underline{optimal region of
integration;$r^{opt}$}. The link integration should be done with
$r^{opt}$\\
\indent
The $r^{opt}$ are shown as a function of $\beta$ 
in Fig.\ref{radius_vs_B}.
Notice that $r^{opt}$ region
changes with the position of link variables on a configuration and 
also with configurations due to the
fluctuation of
gauge fields. The results shown in Fig.\ref{radius_vs_B} are not the
average over the
fluctuations. But the fluctuation of the $r^{opt}$ region is 
not large. If we choose, 
\begin{equation}
  r^{opt} \sim r^{opt}_{mid}= 
\displaystyle{\frac{r^{opt}_{max} + r^{opt}_{min}}{2}}
\end{equation}
it has been in an optimal region of $r$ throughout the link variables and 
configurations.
The $r^{opt}$ for space link $U_{space}$ is
larger than that of $U_{time}$ when $\xi > 1$;
$ r^{opt}_{space} > r^{opt}_{time} $, and 
they become smaller for larger $\beta$. 

In the case of improved actions, the number of link $U$ which are 
simultaneously
integrated in a loop becomes much smaller than the case of standard
action. Therefore the link integration method is not effective for the
calculation of smaller Wilson loops. 
The suppression of the
fluctuation is impressive for $W(6\times6)$ but not for $W(4\times4)$
for $\beta=2.5$.
However for the calculation of
$\eta$ in the smaller $\beta$ region, the use of the link integration
method has been indispensable. We have used it for the calculation of $\eta$
at $\beta=3.5,3.05,2.5$ for Iwasaki action.

\section{Self-energy Effects on $\eta$}
Lattice potential is defined by the ratio of Wilson loops, 
$V(r)=log(\displaystyle{\frac{W(p,r)}{W(p+1,r)}})$.
The $\xi_{R}$ is determined by the matching of the potential in space
and temporal direction.
\begin{equation}
  V_{s}(\xi_{B},r)=V_{t}(\xi_{B},t=\xi_{R} \times r)\label{matching1}
\end{equation}  
The $\eta$ parameter is defined by $\eta=\xi_{R}/\xi_{B}$
However, $V_{s}$ and $V_{t}$ includes self-energy contributions which
may be written as,
\begin{equation}
  V_{s}(\xi_{B},r)=V_{s}^{0}(\xi_{B})+V_{s}(\xi_{B},r)\label{potsigma}
\end{equation}  
Similarly for $V_{t}(\xi_{B},t)$.\\
\underline{ Due to the anisotropy, 
$V_{s}^{0}(\xi_{B}) \neq V_{t}^{0}(\xi_{B})$}.\\
For the standard action, the effect of the self-energy term on $\eta$
has been studied by Bielefeld
group\cite{eta_bielefeld}. It is
reported that the effect on $\eta$ is $\sim 0.01$. In this report
we study its effects for Iwasaki's improved action\cite{Iwasaki}.
In order to get rid of the effect of self-energy term $V^{0}$, we employ a
subtraction method,
\begin{equation}
 V_{s}^{Sub}(\xi_{B},r)=V_{s}(\xi_{B},r)-V_{s}(\xi_{B},r_{0})\label{potsub}
\end{equation}
similarly for $V_{t}^{Sub}$. The $t_{0}$ and $r_{0}$ should satisfy
$t_{0}=\xi_{R}r_{0}$ and we take $r_{0}=3$.
Then we apply the matching for $V^{Sub}$\cite{Klassen}.
\begin{equation}
 V_{s}^{Sub}(\xi_{B},r)=V_{t}^{Sub}(\xi_{B},t=\xi_{R}\times r)\label{etasub}
\end{equation}
We denote the $\eta$ determined in this way as $\eta^{Sub}$. \\
\begin{figure}
\begin{center}
\scalebox{0.30}{ { \includegraphics{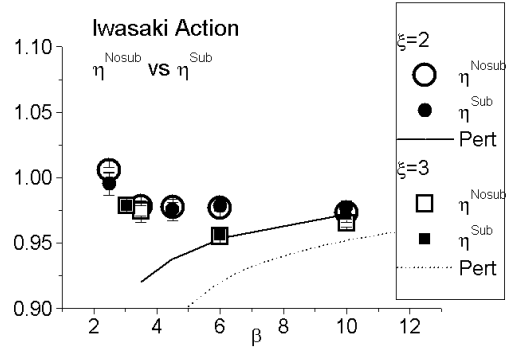} } }
\caption{$\eta^{Sub}$(filled circle) and $\eta^{Nosub}$
without subtraction(open Symbols) at $\xi_{R}=2.0$  \label{eta_sub} }
\end{center}
\end{figure} 
In Fig.\ref{eta_sub} we have shown our results.
It is found that the differences of the $\eta^{Nosub}$ and
$\eta^{Sub}$ are small; $\leq 0.015$.\\
\indent
 Combined with the results of Bielefeld group for the
standard action,  we conclude that the correction to the global 
dependences of
the $\eta$
parameters for the class of improved actions(Symanzik, Iwasaki etc)
which we have presented at lattice 99\cite{sakai2} is really small


\section{Quark Gluon Plasma on Anisotropic Lattice}
As applications of the anisotropic lattice, we have started a
simulation of transport coefficients of the quark
gluon plasma and a study of the heavy quark
spectroscopy\cite{saito}.\\

\begin{figure}
\begin{center}
\scalebox{0.30}{ { \includegraphics{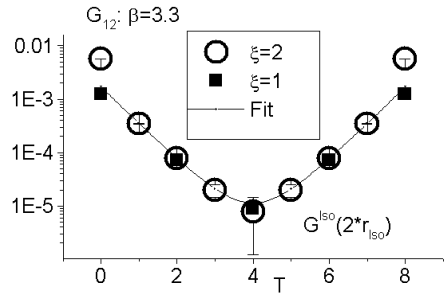} } }
\caption{Matsubara green function $G_{12}$ on $8^3 \times 8$
lattice with $\xi_{B}=2.0$(open circle), and on isotropic $8^3 \times
4$ lattice(filled square) at $\beta=3.3$ \label{g12} }
\end{center}
\end{figure}

We reported the transport coefficients
from lattice simulations on isotropic
lattice\cite{sakai}\cite{sakai1}. 
Our results have been
impressive and encouraging for the phenomenological study of the 
quark gluon plasma, in the sense that they are located close to the
simple
extrapolation of the perturbative results on the figure at high 
temperature limit\footnote{Notice that perturbative formula breaks
down around $T_{c}$}.
However they depend on the ansatz for the spectral
function of the Matsubara green function($G_{\mu\nu}$) of energy
momentum tensor.
The aim of this work is to improve the resolution of the $G_{\mu\nu}$
by using the anisotropic lattice, and check the ansatz
of the spectral function.\\
\indent
The calculation of the transport coefficients of quark gluon plasma
are formulated in the linear response theory. They are calculated by the
$G_{\mu\nu}$ of energy momentum tensor. For the pure gauge
models it is written as,
$ T_{\mu\nu}= 2 \,\mbox{Tr}\, [F_{\mu\sigma}F_{\nu\sigma}
-\frac{1}{4}\delta_{\mu\nu}F_{\rho\sigma}F_{\rho\sigma}]$
On an anisotropic lattice, the field strength tensors are written as
follows,
\begin{equation}
 a_{\sigma}^3 F_{ij}^{c} F_{ij}^{c}  = 
     \frac{\beta_{\xi}}{6 a_{\sigma}} \eta (3-TrU_{ij})
\end{equation} 
\begin{equation}
 a_{\sigma}^3 F_{i4}^{c} F_{i4}^{c}  = 
     \frac{\beta_{\xi}}{6 a_{\sigma}} \xi_{B}^2 \eta (3-TrU_{i4})
\end{equation} 
\indent
We have started from test run of calculating $G_{\mu\nu}$ on 
$8^3 \times 8$
lattice with Iwasaki action at $\beta=3.3$ and $\xi_{R}=2.0$.
As a check of our calculations on the anisotropic 
lattice, we 
have compared it with the one from an isotropic lattice $8^3 \times 4$,
for which the
scale in temperature direction is changed to $\xi_{R}=2$ lattice.
They are shown in Fig.\ref{g12}.  

It is seen that the both green functions coincide with each other.
The merit of applying an anisotropic lattice is that
we could determine the transport
coefficients under the ansatz of the spectral function on small lattices.
\begin{equation}
\begin{array}{ll}  
\rho(\vec{p}=0,\omega) \\
\hspace*{1cm} =\frac{A}{\pi}
(\frac{\gamma}{(m-\omega)^2+\gamma^2}-
\frac{\gamma}{(m+\omega)^2+\gamma^2})\label{ansatz}
\end{array}
\end{equation}
\noindent  
Because 3 independent parameters are determined by the 3 independent
data points of $G_{12}$ on an anisotropic lattice.\\
\indent
We have proceeded to the calculation of Matsubara green function on 
$24^3 \times 16$ lattice with $\xi_{R}=2.0$.  At this point, the
number of data is
not large enough to determine the $G_{\mu\nu}$.
But we expect that the check of the ansatz given by Eq.\ref{ansatz}, 
could be done on the anisotropic lattice. In
addition we are planing to determine spectral functions
by the maximum entropy method.  They will be reported in the
forthcoming publications\\

\noindent
ACKNOWLEDGMENTS\\
This work has been done with SX-5 at RCNP and VX-4 at
Yamagata University. We are grateful for the members of RCNP for kind
supports.\\

\end{document}